\documentstyle[12pt]{article}
\def\om{\omega}

\def\frac#1#2{{\textstyle{{#1}\over {#2}}}}

\def\lsim{\mathrel{\rlap{\lower4pt\hbox{\hskip1pt$\sim$}}
    \raise1pt\hbox{$<$}}}
\def\gsim{\mathrel{\rlap{\lower4pt\hbox{\hskip1pt$\sim$}}
    \raise1pt\hbox{$>$}}}
\def\sqr#1#2{{\vcenter{\vbox{\hrule height.#2pt
         \hbox{\vrule width.#2pt height#1pt \kern#1pt
         \vrule width.#2pt}
         \hrule height.#2pt}}}}

\newcommand{\beq}{\begin{equation}}
\newcommand{\eeq}{\end{equation}}
\newcommand{\bea}{\begin{eqnarray}}
\newcommand{\eea}{\end{eqnarray}}

\renewenvironment{thebibliography}[1]
 { \rm
   \begin{list}{\arabic{enumi}.}
    {\usecounter{enumi} \setlength{\parsep}{0pt}
     \setlength{\itemsep}{3pt} \settowidth{\labelwidth}{#1.}
     \sloppy
    }}{\end{list}}
 
\begin{document}
\titlepage
 
\begin{flushright}
{UWHPE 001\\}
{July 1999\\}
\end{flushright}

\vglue 1cm
	    
\begin{center}
{{\bf Past electron-positron g-2 experiments yielded sharpest bound on CPT violation for "point" particle\\}
\vglue 1.0cm
{H. Dehmelt, R. Mittleman, R. S. Van Dyck, Jr. and P. Schwinberg\\} 
\bigskip
{\it Physics Department, U of Washington\\}
\medskip
{\it Seattle WA 98195, U.S.A.\\}
}
\vglue 0.8cm
 
\end{center}

{\rightskip=3pc\leftskip=3pc
In our experiments  on single $e^-$ and $e^+$ we measured the cyclotron and spin-cyclotron difference frequencies \( \om_c^{\pm}\) and \(\om_a^{\pm} = \om_s^{\pm} - \om_c^{\pm}\) and ratios \( a^{\pm}  = \om_a^{\pm}/ \om_c^{\pm}\) at \(\om_c/{2\pi} \simeq 141 GHz\) and later only $e^-$-values  also at \(\om_c/2\pi \simeq 164 GHz .\) Here we do extract from these data, as has not been done before, a new and very different figure of merit for violation of CPT symmetry, one similar to the widely recognized impressive limit
\[    
{|m_{Kaon} - m_{Antikaon}| / m_{Kaon}  \lsim  10^{-18}}   
\]     
for the K-mesons $\it composed$ of two quarks. That expression may be seen as comparing experimental relativistic mass-energies of particle states before and after the C, P, T operations have transformed particle into antiparticle. Such a similar figure of merit, for a "non composite" and quite different lepton, found by us from our $ \Delta a \equiv a^- - a^+ $ data was even smaller, 
\[ |\hbar (\om^-_a/2 - \om^+_a/2)|/m_0 c^2 = |\Delta a| \hbar \om_c /2m_0 c^2 =|3 \pm  12| \times 10^{-22}
\].}
\vskip 1 cm
PACS: 11.30.Er, 12.20.Fv, 14.60.Cd
\vskip 1 cm
\centerline{\it } 
\begin{center}
{\it Editorially approved for publication in PRL \\}
\end{center}
\medskip
\centerline{\it }
\newpage
\baselineskip=20pt

	In 1962 Dirac wrote: "Recently, new evidence has appeared for the finite size of the electron by the discovery of the muon, having properties so similar to the electron that it may be considered to be merely an excited state of the electron." Finite physical size [1-3] of Dirac's supposed point particles of 1926 may prevent them from obeying [4-6] exact CPT symmetry simply because this theorem has only been proven for mathematical fictions, true point particles. This has revived interest in the sparse CPT violation data currently available and their interpretation. In our experiments [7] on a single $e^- $ and $e^+$ essentially at rest in free space, and later [8] on $e^-$ alone, we measured the cyclotron and spin-cyclotron difference frequencies $\om_c^{\pm}$ and $\om_a^{\pm}$ = $\om_s^{\pm}$ - $\om_c^{\pm}$. Avoiding the well-known problems of obtaining the small difference between two large quantities with non vanishing errors we measured $\om_a^{\pm}$ directly. To this end we induced spin flips by a  spin-flipping rf field at $\om_s^{\pm}$ synthesized from the free electron/positron cyclotron motion at $\om_c^{\pm} = eB/m^{\pm}_0$ and an applied rf field at a precisely measured variable frequency. Spin flips thus produced were observed with the help of  the Continuous Stern-Gerlach Effect [9] and counted. A peak in the counting rate as the applied rf field was swept in frequency over $\om_a^{\pm}$ signaled the resonance. To minimize the effects of unavoidable small drifts of the B-field we recorded the anomaly values $a^{\pm} = \om_a^{\pm}/\om_c^{\pm}$. Our g-value stood for the combination of observed frequency ratios $g^{\pm} = 2(a^{\pm} + 1)$ and we used $ (g^- - g^+)/2 = a^- - a^+$, here now called $\Delta a$, as a measure of CPT symmetry violation, where $g^-, g^+$ denote our g ratios for $e^-$ and $e^+$. 

	To our experiments on "non-composite" leptons Bluhm, Kosteleck\'{y}, and Russell (BKR) [10] have recently applied  their impressive formalism that mildly extends the standard model with the help of small, critically selected CPT breaking perturbations and a perhaps less mild postulated new cosmic axial vector field. In this effort, much appreciated by us, they have been more interested in calling for complex new experimental procedures to test the BKR model than in making full use of our long available data for the purpose of extracting from them the sharpest possible bound on CPT violation. Stimulated by their work, we do here extract from our data, as has not been done before, a new and very different figure of merit, one similar to the widely recognized limit
\beq
|m_{Kaon} - m_{Antikaon}|/m_{Kaon} \lsim 10^{-18}
\eeq
for the K-mesons $\it composed$ of two quarks [11.] Without reference to the BKR model, that expression may be seen as comparing experimental relativistic mass-energies of particle states before and after the C, P, T operations have transformed particle into antiparticle. Accordingly, in the same constant magnetic field we let the symmetry operations transform an electron in the lowest energy level $E^-_{n, s}$, n = 0, s = -1, into a positron in the lowest energy level $E^+_{n, s}, n = 0, s = 1,$ where $n = 0, 1, 2...$  is the cyclotron quantum number and $s = \pm 1$ stands for spin up/down. When CPT symmetry holds we have
\beq
E^-_{0, -1} - E^+_{0, 1} =  (m^-_0 c^2 - \hbar \om^-_a/2) - (m^+_0 c^2 - \hbar\om^+_a/2) = 0
\eeq
or
\beq
E^-_{0, -1} - E^+_{0, 1} - (m^-_0 - m^+_0)c^2 = - \hbar (\om^-_a - \om^+_a)/2 = 0
\eeq
When the right side of the last equation is found not to vanish it becomes a measure of CPT violation and as fraction of $m_0 c^2$ a dimensionless figure of merit for the symmetry violation,
\beq
|\hbar (\om^-_a - \om^+_a)|/2m_0 c^2, 
\eeq
that on first sight appears to require the measurement of $\om^-_a,  \om^+_a$ in more or less exactly the same B field. Actually, this turns out not to be the case.

	Our later experiments [8] have shown that numerical values of $ a^- $ measured at $ \om_c/2\pi \simeq 164 GHz $ differed from those at $ \om_c/2\pi \simeq 141 GHz $  by no more than $1.1 \times 10^{-11}$ compared to the measurement error in $a^{\pm}$ of $\simeq$ $10^{-12}$ and thus the anomaly value is practically constant over a range of 23 GHz. Earlier [7], in the same apparatus with only the trapping potential reversed and therefore in nearly identical fields $B(e^-), B(e^+)$ with cyclotron frequencies for electron and positron differing by less than 56 kHz, we had measured $|a^- - a^+| \lsim 2 \times 10^{-12}$ at $\om_c/2 \pi \simeq 141 GHz$. Since B-field dependent contribution to $a^-$, if any, changed by no more than $1.1 \times 10^{-11}$ over 23 GHz, the effect of not using exactly the same field could have changed the $e^-, e^+$ anomaly values by no more than $3\times 10^{-17}$, a totally negligible amount. Therefore, substituting  $(a^-_{B(e^-)} - a^+_{B(e^+)})\om_c^-  = \Delta a \om_c $ for $(\om_a^- - \om_a^+)$ in equation (4) is quite legitimate and eliminated the need for the feat of making $B(e^-), B(e^+)$ more or less exactly equal proposed later in ref. [10]. Thus we arrive at the final numerical value of the merit figure from our work
\beq
|\hbar(\om^+_a - \om^-_a)|/2m_0c^2 = |\Delta a| \hbar \om_c /2m_0 c^2 = |3 \pm  12| \times 10^{-22}.
\eeq
Our result is now revealed as the sharpest published bound on CPT breaking for a "point" particle. Here and everywhere in the paper our error limits are quoted for one standard deviation.	

	Interpreting our result under the BKR model, the only theoretical model currently available, yields the following expression (6), applying exclusively to CPT violation. Eq. (5) is modified to
\beq
|E^-_{0, -1} - E^+_{0, 1}|/m_0c^2 = |\Delta a| \hbar \om_c /2m_0 c^2 = |3 \pm  12| \times 10^{-22} 
\eeq
as under the BKR model the small CPT symmetry violating perturbations leave electron/positron cyclotron frequencies $\om^{\pm}_c$ and rest masses $m^{\pm}_0$ identical $\om_c$ and $m_0$. While eq. (6) is even more similar to  eq. (1) than eq. (5) it must be mentioned that the assumptions underlying the BKR model and eq. (6) are less general than those underlying eq. (1). Further, we must now address the unpleasant fact that according to the model the conditions of the 1987 experiment [7] may have been less than optimal for detecting CPT violation. According to eq. (8) of ref. [10] the $e^-/e^+$ anomaly frequency splitting is not constant as one might expect naively, but varies with sidereal time 
\beq
(\om_a^- - \om_a^+) = - 4b\hat b \cdot \hat B = - 4b_3
\eeq
where $\vec b$ is a vector of  length $b$ and direction $\hat b$ fixed with respect to the fixed stars but otherwise unspecified. This vector quantifies the degree of CPT violation and thereby becomes the all-important parameter of the model that experiment must determine. $\hat B$ is a unit vector along the magnetic field, which in our lab is directed vertically upward. Obviously $\hat b \cdot \hat B$ changes as the earth rotates around its axis $ \hat e $. To quickly orient ourselves about the range of splittings produced  in our lab located at about latitude $ \pi/4 $ by a given $\vec b$ vector we discuss the 3 special cases of the angle between $ \vec b $ and $ \hat e $ having the values 0, $ \pi/4 $ and $ \pi/2 $. The corresponding value ranges of $\hat b \cdot \hat B$ then are: $const = 1/\sqrt 2 $, $0...1$, $-1/\sqrt 2...0...1/\sqrt 2$. The more or less blind zone where $\hat b \cdot \hat B$ drops below $1/4$ is not negligible, perhaps 30 percent of the whole sphere. It is widest for the angle between $ \vec b $ and $ \hat e $ having the value $ \pi/4$, so we focus on this angle in the following. All 1987 $e^-/e^+$ data were taken in solar not sidereal time over a 42 day period, daily from midnight $\pm$ 1h to 5 AM $\pm$ 1h in which 16 days of $e^+$ data taking was followed by a 5 day pause and then by 21 days of $e^-$ data taking. Each data period is roughly equivalent to one 2:30 AM point, one for $e^+$ at day 8 and one for $e^+$  at day 33. Combined they effectively yield one $\Delta a $ point taken at day 21. The odd AM time slots were necessitated by conditions in our lab. The unfortunate possibility that data taking may have been limited to a partly blind time slot where $|b_3|/b $ had dropped much below $1/4$ may in future work be eliminated by repeating an identical series  $1/4$ $year$ later when it again has reached a large value. The latter occurs because then the earth has completed one quarter of its orbit around the sun and, as seen from our lab,  the stars appear on the sky 6 hours earlier at 8:30 PM where they had been at 2:30 AM on day $1$. We can achieve part of this already with our 1987 data at the price of larger error limits as follows. Combining only the first days of the $e^+$ and $e^-$ data we find  $\Delta a = (- 2.2 \pm 3) \times 10^-12 $ at effectively day 12 while combining only the last days of these data gives  $\Delta a = ( 2.2 \pm 2.2) \times 10^-12$ at effectively day 29. For $ \vec b $ and $ \hat e $  making an angle $ \pi/4 $  this implies that if on day 12 in a worst possible case scenario the ratio $|b_3/b|$ had been 0, 17 days later by day 29 it would have grown to about 5 percent of its peak value $1$. For all possible orientations of $ \vec b $ against $ \hat e $ this value and eq. (6) now allow us to roughly bound b,
\beq
b \lsim 20 radians/sec.
\eeq
By contrast if our data had been taken when the orientation of $\vec b$ was most favorable, namely $\vec B \parallel \vec b$, they would have shown that it must be 
\beq
b \lsim 0.7 radians/sec
\eeq
as it escaped detection. Another result of the BKR model, devastating on first sight, predicts $g^+ = g^-$ when $g$ is interpreted not as ratio of measured frequencies but as correct theoretical gyromagnetic ratio. Obviously our definition $g = 2(a + 1)$ is modified by the CPT violating perturbations here which explains our shift of emphasis from $g$ to $a$ values in the introduction.

\vglue 0.6cm
One of us, H. D., enjoyed discussions with A. Kosteleck\'{y}, D. Boulware and M. Baker. Our colleague I. Ioannou read the manuscript. The National Science Foundation supported this work under Grant 9530678. LateX formating fashioned after reference [10]. 
\vglue 0.6cm

{}

CPT100599.tex

\begin{thebibliography}{}

\bibitem{pd}
Compare, e.g., P. Dirac, Proc. Royal Soc. (London) A268, 57 (1962) 

\bibitem{bp}
B. Pullman, The atom in the history of human thought (Oxford University Press, New York, 1998) 

\bibitem{hd}
H. Dehmelt, Physica Scripta T59, 87 (1995)

\bibitem{kp}
A. Kosteleck\'{y}, R. Potting, Nucl. Phys. B359, 545 (1991) 

\bibitem{ak}
A. Kosteleck\'{y}, CPT Violation, Strings, and Neutral-Meson Systems@, Presented at Orbis Scientiae, January 1997, abstract hep-ph/9704264

\bibitem{jlmn}
J. Ellis, J. Lopez, N. Mavromatos and D. Nanopoulos, Physical Review D 53, 3846 (1996) 

\bibitem{vsd}
R. Van Dyck, Jr., H. Dehmelt and P. Schwinberg, Phys. Rev. Lett. 59, 26 (1987), and Phys. Rev. D 34, 722 (1986).

\bibitem{m}
"Improving the electron g-2 experiment" R. Mittleman, F. Palmer, and H. Dehmelt, in "Traps for antimatter and radioactive nuclei", eds. J. D'Auria, D. Gill and A. Javin (Baltzer, Basel 1993) and Hyp. Int. 81, 105 (1993)

\bibitem{d86}
H. Dehmelt, Proc. Natl. Acad. Sci. USA 83, 2291 (1986). 

\bibitem{bkr}
R. Bluhm,  A. Kosteleck\'{y} and N. Russell, Phys. Rev. Lett. 79, 1432 (1997)
also hep-ph/9707364 and Phys. Rev. D 57, 3932 (1998). 
                                              
\bibitem{c}
C. Caso et al., Euro. Phys. J.  C 3, 1 (1998)


\end{thebibliography}
\end{document}